\documentclass[5p,times]{elsarticle}

\usepackage{textcomp,gensymb}
\usepackage{multirow}
\biboptions{sort&compress}
\usepackage{float}
\usepackage[latin1]{inputenc}   
\usepackage{amsmath} 
\usepackage{epstopdf}           %
\usepackage{flushend}           %
\usepackage{lmodern}
\usepackage[T1]{fontenc}
\usepackage{xcolor}
\usepackage{upgreek}

\usepackage{amssymb}

\usepackage{lineno}

\usepackage[figuresright]{rotating}
\usepackage{color,soul}

\begin{document}
\begin{frontmatter}

\title{3D Mapping of Intragranular Residual Strain and Microstructure in Recrystallized Iron Using 
Dark-Field X-ray Microscopy}

\author[First,Second]{V. Sanna}
\ead{virginia.sanna@esrf.fr}
\author[Third]{Y. Zhang}
\author[First,Second]{W. Ludwig}
\author[First]{A. Shukla}
\author[First]{A. Benhadjira}
\author[First]{M. Sarkis}
\author[First]{C. Yildirim}
\ead{can.yildirim@esrf.fr}
\cortext[cor1]{Corresponding Author}

\address[First]{European Synchrotron Radiation Facility, 71 Avenue des Martyrs, CS40220, 38043 Grenoble Cedex 9, France.}
\address[Second]{Universite Lyon I, INSA Lyon, CNRS, MATEIS UMR5510, 69621 Villeurbanne, France }
\address[Third]{Department of Civil and Mechanical Engineering, Technical University of Denmark, DK-2800, Kgs. Lyngby, Denmark }

\date{\today}

\begin{abstract}
Grain growth is a key process in the thermomechanical treatment of metals. Recently, the presence of local residual stresses within fully recrystallized grains has attracted increasing interest in connection with shear-coupled grain boundary migration mechanisms. In this work, we provide the first direct experimental measurements of residual elastic strain variations in fully recrystallized commercial-purity iron, on the order of $10^{-4}$. Using dark-field X-ray microscopy (DFXM), we performed non-destructive three-dimensional measurements of strain and orientation variations within individual grains. Our results reveal heterogeneous strain distributions across all measured grains. In one case, we observed several isolated dislocations accommodating two second-phase particles, exhibiting a localized strain signature with no detectable long-range effect. The formation mechanisms of intragranular residual strains and their potential influence on grain boundary migration during subsequent grain growth are discussed. 
This work highlights the importance of accounting for such residual elastic strains in future grain growth models.
\end{abstract}


\end{frontmatter}

\begin{sloppypar}

In thermomechanical processing, metals are plastically deformed and then heat-treated to tailor their mechanical properties. After plastic deformation, such as cold rolling, hardness increases while ductility decreases. Annealing restores ductility and relieves internal stresses through recrystallization, which forms new grains generally assumed to be nearly defect- and stress-free \cite{HUMPHREYS2004}. Consequently, models of grain growth, such as analytical approaches \cite{Srolovitz, HILLERT}, phase-field \cite{zhang2024}, Monte Carlo \cite{Zollner2007} and level-set approaches \cite{BERNACKI}, often neglect residual stress in recrystallized microstructures. Only few studies \cite{zhang2024,CHEN2019} have examined boundary migration mechanisms beyond diffusion, accounting for the fact that diffusion alone cannot fully explain boundary motion \cite{KANG2023,Han2018,Thomas2017}.

However, even small residual strains within recrystallized grains can influence subsequent microstructural evolution. Bicrystal experiments have demonstrated that grain boundary motion is sensitive to strains as low as $\sim10^{-6}$ \cite{Gorkaya2009, MOLODOV2007, SHARON2011, MOMPIOU2009}, also enforced by in-situ Transmission Electron Microscopy (TEM) observations of stress-assisted grain growth in nanocrystalline Al \cite{Legros2008}, suggesting that residual elastic strain may affect grain growth kinetics in polycrystalline materials. Understanding the presence and role of such strains is therefore essential for accurately describing grain growth behavior. 

Experimental evidence of local residual strain has been reported mainly in partially recrystallized metals, including aluminum \cite{AHL2017}, nickel \cite{Zhang2019}, iron \cite{ZHANG2022}, and gum metal \cite{Lindkvist2023}, with strain magnitudes on the order of $10^{-4},10^{-3}$. These observations were enabled by synchrotron-based X-ray methods that offer bulk, non-destructive strain mapping, such as differential aperture X-ray microscopy (DAXM) \cite{Larson2002_3DXSM} and scanning 3DXRD \cite{Henningsson2024}. In contrast, electron microscopy techniques such as Transmission Electron Microscopy and Electron Back Scatter Diffraction (EBSD) can also detect local strain \cite{RUGGLES2026,Kassner2002,JIANG2013,KASSNER2013}, though are limited to near-surface regions, affected by surface relaxation, and typically achieve a resolution of only about $10^{-4}$. Once the deformation matrix has completely disappeared, as in a fully recrystallized sample, residual strain levels are expected to be even lower, posing a significant technical challenge for accurate characterization. 


Techniques such as Bragg coherent diffraction imaging (BCDI) offer nanometer-scale spatial resolution and can directly image lattice defects such as dislocations and twin boundaries \cite{Newton2010, Clark2013, Ulvestad2015}, but the need for high beam coherence and non-complex crystals limits its use to polycrystalline bulk samples with crystal thicknesses of only 1 to 2 $\micro$m. Similarly, X-ray Topography can reach small strain resolution, but requires the sample to have grains of the order of at least 100 $\micro$m or to be a single crystal \cite{topo}. Consequently, no three-dimensional measurements have yet resolved intragranular residual strain fields in \textit{fully} recrystallized grains due to these experimental limitations.

Dark-field X-ray microscopy (DFXM) \cite{Simons2015, Poulsen2017, Yildirim2020} bridges this gap by combining non-destructive, 3D bulk characterization with sensitivity to intragranular strain fields on the order of $10^{-5}$, and an angular and spatial resolution of $0.001\degree$ and 100 nm, respectively \cite{Poulsen2018bfp}. Using an objective lens along the path of the diffracted beam, DFXM enables direct imaging of lattice distortions, orientation gradients, and strain variations within individual grains. Therefore, it provides a unique opportunity to study residual elastic strain and defect structures in fully recrystallized materials. Here, we employ DFXM to investigate intragranular strain and orientation fields, along with the associated defect structures, within recrystallized grains of polycrystalline pure iron.

\begin{figure}[htbp]
  \includegraphics[width=0.8\linewidth]{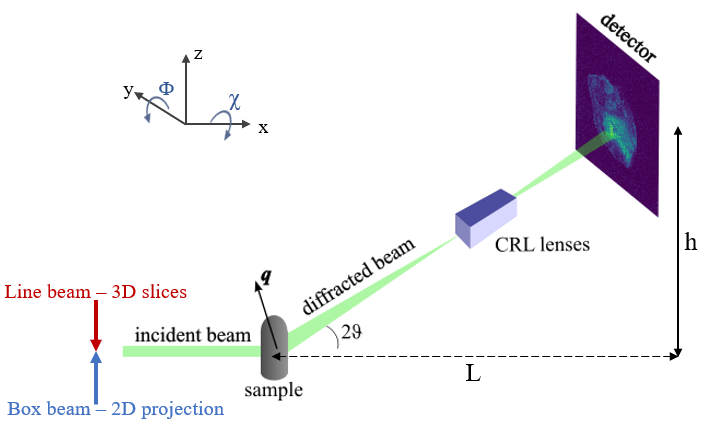}
  \caption{ Schematics of the DFXM setup showing the incident beam illuminating the sample and the diffracted beam passing through the X-ray objective to form a dark field image with diffraction contrast sensitive to lattice spacing and misorientation along the 110 vector of iron. Sample rotations $\phi$ and $\chi$ are indicated. Two illumination modes were used: (i) a box beam (blue) providing full-volume projection of the diffracting grain and (ii) a line-focused beam (red) provides information through the height of the grains, as the sample is translated across the horizontal beam in the $z$ direction to obtain three-dimensional information, enabling height-resolved imaging with a spatial sensitivity of about $500\,\mathrm{nm}$ in this direction.}

  \label{fig:setup}
\end{figure}

In this study, we examined a fully recrystallized commercial purity iron ($99.9\,\%$). To achieve this state, the material was first cold rolled to a $90\,\%$ thickness reduction and then annealed at $700\,^\circ\mathrm{C}$ for 30\,min. A needle-like sample was cut from the rolled plate by Electrical Discharge Machining (EDM) and its surface was subsequently electro-polished to remove any damage introduced during the cutting. The specimen measured approximately $8\,\mathrm{mm}$ in length, with thickness varying along its height as shown in Fig.~\ref{fig:setup}. DFXM measurements were performed near the top section of the sample, within a local cross-sectional volume of a cylinder with a $200\,\mu\mathrm{m}$ diameter. The average grain size in this region was 15--20\,$\mu\mathrm{m}$, as determined by light optical microscopy (LOM).

DFXM experiments were carried out at two beamlines of the European Synchrotron Radiation Facility (ESRF): the former ID06-HXM \cite{Kutsal2019} and the upgraded EBS beamline ID03 \cite{isern2025}. Both beamlines employed the same diffraction geometry and measurement principles, with improvements in beam conditioning and optical stability at ID03. A monochromatic incident beam was selected using the (111) reflection from a silicon channel cut crystal, yielding a relative bandwidth of $\Delta E / E = 10^{-4}$. A vertical diffraction geometry was used to probe the 110 reflection of body-centered cubic (BCC) iron, with a nominal $d$-spacing of 2.026~\AA  \cite{dspacing}. A total of seven individual grains were studied using two illumination modes: (i) a box-shaped parallel beam with a $200 \times 200\,\mu\mathrm{m}^2$ cross section, and (ii) a focused line beam produced using 1D compound refractive lenses (CRLs) made of beryllium with $100\,\mu$m radius, placed $720\,\mathrm{mm}$ upstream of the sample. In the second mode, 3D data were acquired by translating the sample across the sheet-shaped beam ($200\,\mu$m wide and $500\,\mathrm{nm}$ tall) in $1\,\mu$m steps. Specifically, the line beam mode was used for G1, G2, G3 and G7 and the box beam mode with projection measurements for G4, G5 and G6. In the latter case, the grain shapes were approximated as spherical, in order to enable a direct comparison with the layer-resolved measurements. 

The diffracted beam from a given grain was collected using an X-ray objective composed of 87 2D Be CRLs, each with an apex radius of $50\,\mu\mathrm{m}$, aligned along the diffracted beam direction. The objective was positioned approximately $260\,\mathrm{mm}$ downstream of the sample, while the far-field detector was located at a distance of $\approx 5340\,\mathrm{mm}$ from the sample. This optical configuration resulted in an X-ray optical magnification of approximately $\approx - 18.3\times$ obtained from the CRL objective. The magnified image was projected onto a high-resolution detector system consisting of a scintillator, mirror, and visible-light objective, yielding an effective pixel size of approximately $35\,\mathrm{nm}$ on the detector. Components of the crystallographic orientation variation were measured by the so-called mosaicity scans, which probes the $\phi$ and $\chi$ angles corresponding to rotations about the sample $y$ and $x$ axes, respectively, as illustrated in Fig.\ref{fig:setup}. These mosaicity scans were performed at a fixed scattering angle ($2\theta$). The typical $\phi$ scan ranged from $0.02^\circ$ to $0.06^\circ$, using a step size of $0.002^\circ$, while the $\chi$ angle was scanned over $0.2^\circ$ using a step size of $0.02\deg$.

Strain scans were performed by scanning a 2D mesh of $\phi$ and $2\theta$, enabling the measurement of elastic strain normal to the lattice planes of interest. The strain $\epsilon$ was calculated from the angular shift of the Bragg peak using the relation \cite{GUSTAFSON2023}:
\[
\epsilon = -\frac{1}{2} \frac{\Delta 2\theta}{\tan(\theta_{0})}
\]
The $2\theta_{0}$ angle was calculated as
\[
2\theta_{0} = \arctan\left(\frac{h}{L}\right),
\]
where $h$ is the detector height relative to the direct beam and $L$ is the sample-detector distance. The resulting $2\theta$ values were $20.72^\circ \pm 0.003^\circ$. The total $2\theta$ scan range was approximately $0.02^\circ$, with a typical step size of $0.005^\circ$.

Note that sub-step angular resolution was obtained in the generated maps by fitting the intensity profiles with a Gaussian function. Data processing was performed using the open-source \textit{darfix} package \cite{GarrigaFerrer_darfix}, developed specifically for the treatment of DFXM data. It is important to emphasize that, at the very small intragranular misorientation and elastic strain levels resolved in this study, geometrical corrections in DFXM measurements become critical. In contrast to deformed grains\cite{YILDIRIM2022}, where such effects may be negligible relative to large orientation spreads, here even minor position-dependent shifts can bias the extracted center-of-mass values. The measured $\chi$ and $2\theta$ center-of-mass maps were therefore rigorously corrected using the geometrical relations described in \cite{Poulsen2017, ahl2018elements}, which account for the correlation between the position of the diffracting point in the illuminated plane and the effective angle of incidence on the objective. Corrections equivalent to Eq.27-29 in Poulsen 2017 \cite{Poulsen2017} were applied to remove position-induced shifts due to the coupling of the real and reciprocal space in the objective's lenses in both orientation and strain measurements. Importantly, these corrections were implemented directly from the analytical expressions rather than approximated by empirical trend functions, ensuring accurate quantification of the small residual strain fields reported here. See the section \textit{Geometrical corrections of center-of-mass maps} in Supplementary Materials for details.

\par

Out of the seven grains analyzed, we first focus on a representative grain, denoted G2, with a size of 40 $\micro\mathrm{m}$, to investigate its internal structure in detail. 

\begin{figure}[h!]
    \centering
    \includegraphics[width=1\linewidth]{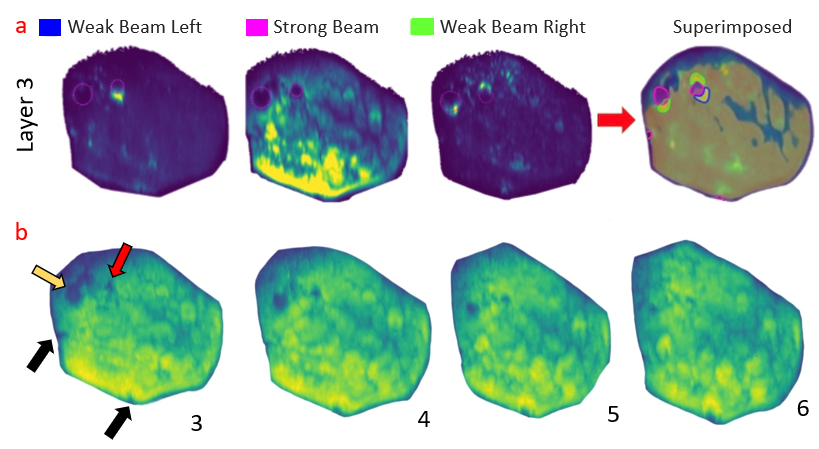}
    \caption{\textbf{(a)} Weak-beam condition (left and right), highlighting dislocations, and strong-beam condition, showing two regions of reduced contrast that likely correspond to second-phase particles (highlighted in pink). Following the red arrows, a superimposed image of the three conditions shows the positions of the particles (pink) and the dislocation contributions from the left (blue) and right (green) weak-beam conditions. \textbf{(b)} maximum value of the rocking curve (peak of intensity during the mosaicity scan) of the layers 3, 4, 5, and 6, respectively. The arrows indicate the position of the particles shown in Fig.\ref{fig:strain_mosa}.}

    \label{fig:comparison}
\end{figure}

Focusing on a single layer of grain G2, Fig.~\ref{fig:comparison}(a) shows the raw data of layer 3 under the strong-beam condition (center) and the weak-beam conditions (left and right). The weak beam condition is the region farthest away from the diffraction peak where contrast is still present, allowing improved visualization of local orientation variations caused by e.g. dislocations \cite{Jakobsen, yildirim2023extensive, abdou2025}. The weak signals in both the weak and strong beam conditions in the region marked by the pink circles are therefore likely second-phase particles, see Fig. S1 in the Supplementary Materials. Clear orientation variations are visible on opposite sides of these particles, as highlighted in the superimposed image, where the regions with large local orientation variation around the particles are shown in blue for the left and in green for the right weak-beam conditions. 

These local orientation variations are likely accommodated by dislocations, which are generated during cooling to relax local thermal stresses arising from the different thermal expansion coefficients of the particles and the iron matrix \cite{Zhang2019}. Furthermore, the orientation variations observed around the two particles exhibit different $2\theta$ dependencies. For example, the bright region below the left particle appears under the right weak-beam condition, whereas for the right particle it appears under the left weak-beam condition. This suggests different dislocation configurations around the two particles.
Fig.~\ref{fig:comparison}(b) shows the raw diffraction images of the maximum diffraction-peak intensity for layers 3, 4, 5, and 6. The yellow, red, and black arrows highlight regions of local minima of diffracted intensity. By following the same regions in the subsequent layers, the evolution of the particles' shape and size can be tracked, showing a progressive reduction of their contrast and extent for the grain interior particles. 
This suggests that these particles gradually become less detectable as the probed volume moves deeper into the grain, ending around layer 6.

\begin{figure*}[htbp]
    \centering
    \includegraphics[width=0.9\linewidth]{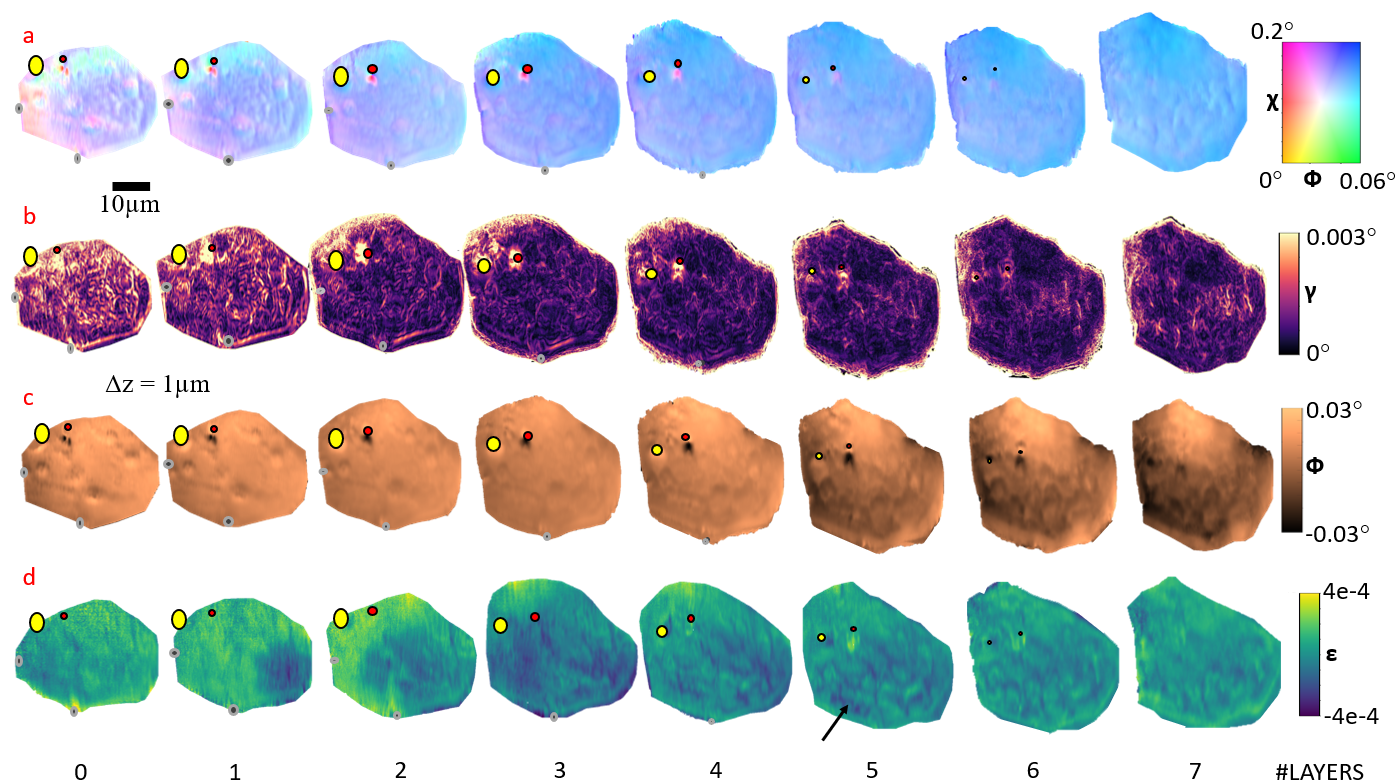}
    \caption{DFXM results on eight layers ($0-7$) of a single grain of interest (G2). The layers are separated by $1\micro$m in height. The different rows represent \textbf{(a)} mosaicity, \textbf{(b)} kernel average misorientation, \textbf{(c)} local orientation maps of $\phi$. \textbf{(d)} strain scans. Next to each scan, there is the corresponding colormap. The grain is approximately $\sim40\micro$m large and contains some particles, represented by circles, one red and one yellow with a black outline and other two in black and grey, and several dislocations surrounding them.}
    \label{fig:strain_mosa}
\end{figure*}

Fig.~\ref{fig:strain_mosa} shows the local 3D orientation and strain variations across several $1~\micro\mathrm{m}$-layers of grain G2 embedded in the sample by using a line-focused beam. 
As expected, the grain shape evolves along its height, i.e. different 2D layers, and the signal remains clear across layers.

Panel (a) provides information on lattice rotation. The mosaicity maps show the lattice tilts and the misalignment with respect to the 110 diffraction vector, essentially representing a radial center-of-mass (COM) map color-coded like a local pole figure around a (110) pole of the grain. 
A pronounced orientation gradient extending from the first to the last layer of the grain is observed, particularly across the first three layers. 
Locally, two closely-spaced regions with large orientation gradients extend across most layers. As previously discussed, these features correspond to accumulations of dislocations surrounding second-phase particles. These particles are highlighted by ellipses/circles outlined in black: one, indicated in yellow, and the other, exhibiting much brighter contrast, in red.
Two additional particles are observed near the grain boundary, and their positions are indicated by grey circles. In the latter cases, no clear agglomeration of dislocations or significant local orientation change is detected. One possible explanation is that the DFXM signal decreases in intensity near the grain boundary, reducing the sensitivity to strain and orientation variations in this region.

The dislocations are visible also in the kernel average misorientation (KAM) map in panel (b). The quantity $\gamma = \sqrt{\Delta \chi^{2} + \Delta \phi^{2}}$ captures the deviation from the local orientation changes between neighboring voxels within each layer \cite{YILDIRIM2022}, highlighting local distortions of the lattice orientation field near the dislocation sites (e.g. surrounding the yellow and red particles).
Panel (c) presents the local orientation variation (lattice tilt) map of the angle $\phi$, chosen for its higher angular sensitivity compared to $\chi$ \cite{Poulsen2017,poulsen2021geometrical}. Most of the layers exhibit a narrow angular distribution below $0.05^\circ$, but distinct local contrasts are visible near the red particle.

The axial strain distribution along the [110] direction is shown in panel (d), with a total spread of $\sim 8\times10^{-4}$.
The strain is non-uniform, with localized fields around the particles and near grain boundaries. Minor fluctuations, visible near the lower boundary of layer 5 for example (see black arrow), are likely attributed to dynamical diffraction effects from multiple scattering through the sample thickness, which are spatially confined and do not affect the overall strain spread. 



The particles are present across layers 0-6, but continue to affect the local orientation variations in layer 7. While the presence of second-phase particles can introduce local dislocations in their vicinity, they can also pin dislocation motion during annealing. This pinning may lead to a higher dislocation density on the left side of the grain, and thus to the relatively large orientation gradient observed in the first layers in Fig.~\ref{fig:strain_mosa}a.


The sharp orientation change adjacent to the red particle persists up to layer 5 and remains visible in layer 6, where the particle appears to terminate. It should be recalled that the measured strain is sensitive only to the lattice distortion component projected along the scattering vector; therefore, under the present imaging conditions, the contrast is expected to arise predominantly from the top and bottom regions of the particle. These features could correspond either to the lattice mismatch or directly the agglomerations of dislocations in this region, both induced by the presence of the particle; however, they produce a local strain response that is mainly detected near the particle tip, consistent with this directional sensitivity.

The yellow particle, although associated with less pronounced orientation changes, also exhibits a localized strain response, consistent with the sensitivity discussed above. Both cases are shown in Fig. \ref{fig:strain_mosa}(d), with values significantly above the average background. The regions adjacent to the red and yellow particles display predominantly tensile and compressive components, respectively. Since both features are located near the particle terminations, they are likely related to the interaction between the particle and the surrounding grain lattice.

\begin{figure}[t]
  \includegraphics[width=0.98\linewidth]{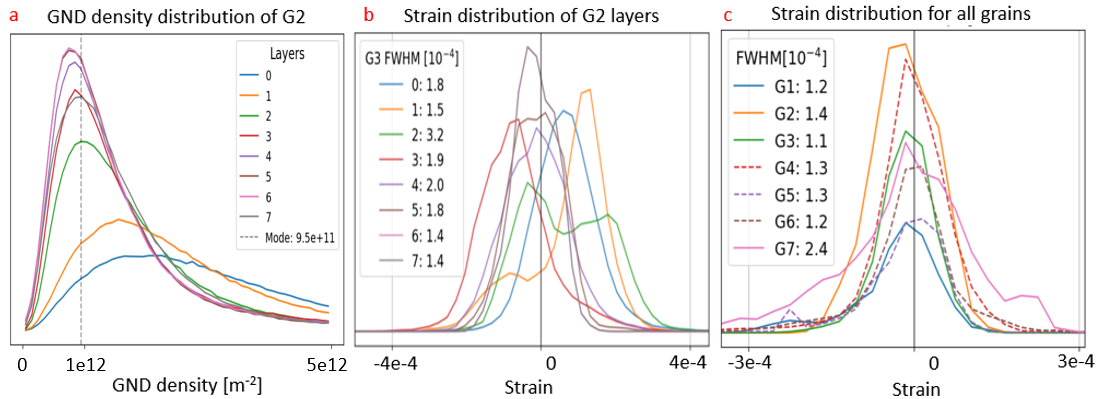}
  \caption{\textbf{(a)} Dislocation density distributions of the same layers of grain G2 as shown in Fig.~\ref{fig:strain_mosa}. \textbf{(b)} Area-weighted strain distributions of the individual layers of grain G2. \textbf{(c)} Volume-weighted strain distributions of all investigated grains. In (c), solid lines correspond to grains for which multiple layers were measured using the line beam, while dashed lines correspond to strain distributions obtained from projection measurements using the box beam; for these grains, a spherical grain shape was assumed to estimate the contributing volume.
}
  \label{fig:g2}
\end{figure}

In Fig.~\ref{fig:g2}(a) and (b), the dislocation density distributions and the area-weighted strain distributions of all layers of grain G2 are presented, respectively, corresponding to the layers shown in Fig.~\ref{fig:strain_mosa}. The geometrically necessary dislocation (GND) density shown in Fig. \ref{fig:g2}(c) was calculated using the following equation from \cite{moussa2015,pho}: 

\[
\rho_{GND} = \frac{\alpha \gamma}{\textit{b} x}
\]

where $\gamma$ is the KAM misorientation angle, $\alpha$ is a constant (here considered 1), $b$ is the Burgers vector, and $x$ is the unit length (pixel size).

Two layers, $0$ and $1$, exhibit a larger GND density spread, as shown in Fig. \ref{fig:g2}(a), corresponding to the same layers in which the orientation maps showed a more prominent change. The presence of the particles, as well as possible interactions with neighboring grains, may reflect the relatively high dislocation density observed in these layers. 

 In general, the values are of the order of $10^{12}\text{m}^{-2}$, which is slightly higher than expected for a fully recrystallized pure sample, presumably due to the presence of particles, but still consistent with the results reported in \cite{ZHANG2022}. When the particles are no longer present, as in layer 7, the distribution is much more concentrated in lower density values and with a smaller spread.

Fig.~\ref{fig:g2}(b) shows the contribution of the individual layers to the overall strain of G2. Layers 0 and 1 exhibit a peak shifted toward the positive side, which is reflected in the strain map in Fig.~\ref{fig:strain_mosa}, where there is a clear predominance of tensile strain and only a compressive region that becomes larger as we move through the layers. Moving up to layer 3, the compressive strain becomes predominant, with only a localized tensile region remaining close to the upper boundary. This behavior may also be influenced by the presence of a neighboring grain. It should also be noted that the interlayer variation is on the same order as the intralayer variation.


Despite the presence of particles in G2, the overall elastic strain distribution does not show any major difference compared with that of the other grains that do not contain particles, irrespective of scanning mode. Fig.~\ref{fig:g2}(c) shows the volume-weighted strain distributions of all investigated grains (G1 to G7) and their respective FWHM values in the legend. In the Supplementary Material, the section \textit{Results of another grain} shows the analysis of grain G3 with no particles.

Across all grains, the FWHM of the strain distribution is $1.4 \times 10^{-4}$ (individual grain values are given in the legend of the strain distribution plot). As expected, these are lower than those reported for recrystallizing grains in partially recrystallized iron from the same batch of samples \cite{ZHANG2022}. 
Importantly, these residual strains are resolved here for the first time. To our knowledge, no existing experimental technique offers strain sensitivity comparable to DFXM at this length scale, making this the first direct experimental observation of local residual strain in fully recrystallized iron.

Overall, the results show that the internal structure within individual grains in this fully recrystallized sample is not uniform: discrete dislocations (see Dislocation Contrast section in Supplementary materials), elastic strain heterogeneity, and lattice rotation gradients all contribute to a complex microstructure. The presence of particles certainly contributes to the local strain and orientation variations; however, their impact is typically confined to the layers in which they are present. Moreover, the two particles seen within G2 are the only ones observed among the seven inspected grains, reflecting their low volume fraction of approx. 0.04\%, as revealed by phase-contrast tomography (see Fig. S1 in the Supplementary Material). The influence of these second-phase particles is therefore considered less critical for the overall strain and defect distribution in the fully recrystallized pure metals reported here, although their presence should be considered when assessing local boundary migration around them in dynamic studies. 

The local residual strain does not show a systematic correlation with intragranular orientation, indicating that this parameter alone is insufficient to explain the observed strain fields. Consequently, geometrically necessary dislocations are unlikely to be the dominant contributors to the measured residual stress. The presence of particles appears to affect the strain state only locally, inducing either tensile or compressive strain in their immediate vicinity. In contrast, interactions with neighboring grains likely play a more significant role in the development of residual stress, as during grain growth, grains can exert mechanical forces to each other, generating regions of non-negligible stress even in the absence of pronounced intragranular defects \cite{yildirim20253d}.

It has already been shown that grain boundary velocity is not only controlled by curvature \cite{BIZANA2023, xu2024grain,bhattacharya2021}, in contrast to what classical theory assumes, but it may be affected by shear and elastic strain \cite{Thomas2017,Qiu2025}. Recent in-situ pink-beam DFXM measurements \cite{yildirim20253d} further revealed that such lattice distortions can accumulate even during growth of fully recrystallized grains, which may lead to residual strains that persist and affect subsequent grain boundary motion through enhanced mechanical interactions with neighboring grains. Therefore, the role of such local residual strains need to be clarified and incorporated into advanced models for grain growth \cite{Chen2002,Steinbach2009}.



In conclusion, we employed DFXM to analyze seven individual grains in fully recrystallized commercial pure iron, revealing three-dimensional strain heterogeneity persisting within nominally strain-free recrystallized microstructures. Enabled by the exceptional strain sensitivity of the technique, intragranular residual strain variation on the order of $10^{-4}$ and dislocation structures were resolved at the sub-micrometer scale. 
These results challenge the classical assumption that such grains are stress-free and highlight the need to incorporate internal stresses into models of grain growth. 

Building on recent advances in targeted grain selection and the seamless integration of 3DXRD/DCT with DFXM \cite{shukla2025} and automated machine-learning approaches trained within the geometrical-optics framework, ongoing and future work could 
map the full strain tensor, and perform in-situ annealing studies to investigate how residual strain fields interact across grain boundaries to influence grain boundary mobility. 
These developments provide direct experimental access to the mechanisms governing grain growth evolution and enable validation of their models in bulk polycrystalline samples, representing a step toward predictive, microstructure-informed metallurgy.

\section*{Acknowledgements}
We thank ESRF for providing the beamtime at ID06-HXM and ID03. CY, AS, and VS acknowledge the financial support from the ERC Starting Grant nr 10116911. YZ thanks Villum Fonden (grant No. VIL54495 MicroAM). The authors gratefully acknowledge Carsten Detlefs for their valuable assistance during the experiments.

\par

\bibliography{ref_rx}

\end{sloppypar}
\end{document}